\documentclass[
a4paper,%
12pt,%
]{article}

\usepackage[dvips]{graphicx}

\title{Method of multiple internal reflections in description of
tunneling evolution through barriers}

\author{
Vladislav~S.~Olkhovsky\thanks{E-mail: olkhovsk@kinr.kiev.ua} \, and
Sergei~P.~Maydanyuk\thanks{E-mail: maidan@kinr.kiev.ua} \\
\small\emph{Institute for Nuclear Research,
National Academy of Sciences of Ukraine,} \\
\small\emph{prosp. Nauki, 47, Kiev-28, 03680, Ukraine}}

\date{\today}

\begin{document}
\begin{sloppypar}
\maketitle

\begin{abstract}
A method of a non-stationary description of tunneling of a particle
through the one-dimensional and spherically symmetric rectangular
barriers on the basis of analisis of multiple internal reflections
of wave packets in relation on the barrier boundaries, named as
\emph{the method of multiple internal reflections}, is presented
at the first time.
For the one-dimensional problem the applicability of this method
is proved, its specific features are analyzed.
For the spherically symmetric problem the amplitudes of the
transmitted and reflected wave packets, times of tunneling and
reflection in relation to the barrier are calculated using this
method.
The effect of Hartman-Fletcher is analyzed.
\end{abstract}


{\bf UDC 539.14}



\section{Introduction
\label{sec.0}}


As the further development of the time analysis of tunneling
processes considered in
\cite{Olkhovsky.1997.JPIF,Olkhovsky.1992.PRPLC,Olkhovsky.1997.Trieste},
here we present the non-stationary solution method for the problem
of scattering of nonrelativistic particle on the spherically symmetric
field which has rectangular barrier. Such approach uses the account
of multiple internal reflections of fluxes in relation to the boundaries
of the barrier (this approach further we name as \emph{the method of
multiple internal reflections} and we present such a method at the
first time).

An important specific feature of this method is a consideration
of particle propagation on the basis of wave packets (WPs).
Due to this one can fulfill the time analysis of particle propagation
(or tunneling) and study in details these processes in interesting time
moment or in relation to the concrete point of space.
In result, the validation of method on the basis of the account of the
multiple internal reflections of WP is given correctly both for
above-barrier and sub-barrier regions (unlike stationary consideration
of such approach to the problem solution
\cite{McVoy.1967.RMP,Anderson.1989.AJPIA}).

At the beggining we consider the problem of tunneling of particle
through an one-dimensional rectangular barrier.
This problem is test one
and allows to analyze the specific features of this method.
The time analysis of the method of multiple internal reflections
is fulfilled for the first time.

The interesting perspective of this method is enough easy calculation
(in comparison with the known stationary methods) of stationary
wave functions (WFs) for the one-dimensional
and spherically symmetric problems with barriers of more composite
form than rectangular. It is shown weakly in the one-dimensional problem
with one rectangular barrier and is shown more brightly in the problems
with barriers of more composite forms. As an example we consider the
one-dimensional problem with the double-humb rectangular barrier.

Further the method is used for solving the problem of scattering
of particle on the spherically symmetric field, which radial part has
a rectangular barrier. For it the amplitudes of transmitted and
reflected WPs, total times of tunneling and reflection in relation to
the barrier are found. The effect of Hartman - Fletcher is considered.
The expression for the S-matrix is presented in the form of the sum of
two components corresponding to the amplitudes of transmitted and
reflected WPs. Spherically symmetric problem with use of the method of
multiple internal reflections is solved for the first time.

\section{The tunneling evolution of particle through the
1D- rectangular barrier}

\begin{figure}[p]
\centerline{\includegraphics[width=7cm]{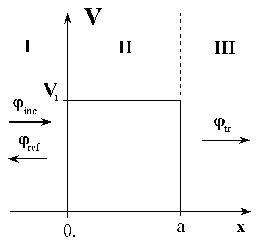}}
\caption{An one-dimension one-humb rectangular barrier}
\label{fig.1}
\end{figure}

Let's consider the problem of particle moving along the $x$-direction
and tunneling through a rectangular potential barrier (see
Fig.~\ref{fig.1}).
We label the region I for $x < 0$, region II for $0 < x < a$ and
region III for $x > a$, accordingly. Solving the stationary
Schr\"{o}dinger equations, one can find the general solution for WF:
\begin{equation}
\varphi(x) = \left \{
\begin{array}{ll}
   e^{ikx} + A_{R} e^{-ikx},            & \mbox{for } x < 0; \\
   \alpha e^{-\xi x} + \beta e^{\xi x}, & \mbox{for } 0 < x < a; \\
   A_{T} e^{ikx},                       & \mbox{for } x > a;
\end{array} \right.
\end{equation}                                          
where $k = \frac{1}{\hbar} \sqrt{2mE}$,
    $\xi = \frac{1}{\hbar} \sqrt {2m (V_{1} - E)}$,
    $E$ and $m$ are the energy and mass of a particle, accordingly.
The coefficients $A_{T}$, $A_{R}$, $\alpha$ and $\beta$ can be
calculated analytically, using requirements of the continuity of WF
$\varphi(x)$ and its derivative on each boundaries of the barrier.

The tunneling evolution of particle can be described using
non-stationary consideration of the propagating WP constructed on
the basis of solution of the stationary Schr\"{o}dinger equations:
\begin {equation}
  \psi(x, t) = \int\limits_{0}^{+\infty} g(E - \bar{E}) * \varphi(x, k)
                e^{-iEt/\hbar} dE,
\end{equation}                                  
where the weight amplitude $g(E - \bar{E})$ satisfies to the requirement
of the normalization $\int |g(E - \bar{E})|^{2} dE = 1$, value $\bar{E}$
is the average energy.

By inserting in the integral (2) instead of the total wave $\varphi(x, k)$
the incident $\varphi_{inc}(x, k)$, transmitted $\varphi_{tr}(x, k)$ or
reflected part of WF $\varphi_{ref}(x, k)$, defined by expression (1),
we obtain the incident, transmitted or reflected WP, respectively.
Considering only sub-barrier processes, we exclude the component of waves
for above-barrier energies, including the additional transformation:
\begin{equation}
   g(E - \bar{E}) \to G(E - \bar{E}) =
   g(E - \bar{E}) \theta (V_{1} - E).
\end{equation}                                  


The method of multiple internal reflections
considers the process of WP propagation sequentially on steps
of its propagation in relation to each boundary of the barrier.

At the first step we consider WP in region I, which is incident
upon the first (initial) boundary of the barrier. Let's assume, that
this packet transforms into the WP, transmitted through this boundary
and tunneling further in region II, and into the WP, reflected from
boundary and propagating back in region I. Thus from the reasons of
elementary causality we consider, that WP, tunneling in region II,
is not reached the second (final) boundary of the barrier
because of a terminating velocity of its propagation, and consequently
at this step we consider only two regions I and II. Because of the same
physical reasons to write the expression for this packet, we consider,
that its amplitude should decrease in positive $x$-direction. For this
we use only one item $\beta*\exp(-\xi x)$ in expression (1), throwing
the second increasing item $\alpha*\exp(\xi x)$ (in the opposite case
we break the requirement of the finiteness of WF for an indefinitely
wide barrier). In result, for the incident, transmitted and reflected
WPs in relation to the first boundary one can write:
\begin{equation}
\begin{array}{lcll}
\psi^{1}_{inc}(x, t) & = & \int\limits_{0}^{+ \infty}
        G(E - \bar{E}) * e^{ikx -iEt/\hbar}dE,
        & x < 0, \\
\psi^{1}_{tr}(x, t)  & = & \int\limits_{0}^{+ \infty}
        G(E - \bar {E}) * \beta^{0} e^{-\xi x -iEt/\hbar} dE
        & 0 < x < a, \\
\psi^{1}_{ref}(x, t) & = & \int\limits_{0}^{+ \infty}
        G(E - \bar{E}) * A_{R}^{0}e^{-ikx -iEt/\hbar}dE,
        & x < 0.
\end{array}
\end{equation}                                          


We assume, that during a particular interval of time considered
at the first step, the transmitted WP tunnels in the sub-barrier
region in the positive $x$-direction. In result, at this step we
receive the flux for this packet, not equal to zero. Considering
only the stationary expressions for WF for this step, we receive
the zero fluxes in the barrier region
\cite{McVoy.1967.RMP,Anderson.1989.AJPIA}.

Constructing from (4) total WF in regions I and II and using
the condition of continuity of this WF and its derivative,
we find unknown coefficients $\beta^{0}$ and $A_{R}^{0}$.

At the second step we consider WP, tunneling in region II and
incident upon the second boundary of the barrier at point $x = a$.
It transforms into the WP, transmitted through this boundary and
propagated in region III, and into the WP, reflected from boundary
and tunneled back in region II. Let's define them so:
\begin{equation}
\begin{array}{lcll}
\psi_{inc}^{2}(x, t) & = & \psi_{tr}^{1}(x, t),         &
        \ 0 < x < a,                                    \\
\psi_{tr}^{2}(x, t) & = & \int\limits_{0}^{+ \infty}
        G(E - \bar{E}) * A_{T}^{0}e^{ikx -iEt/\hbar}dE,  &
        \ x > a, \\
\psi_{ref}^{2}(x, t) & = & \int\limits_{0}^{+ \infty}
        G(E - \bar{E}) * \alpha^{0}e^{\xi x -iEt/\hbar}dE, &
        \ 0 < x < a.
\end{array}
\end{equation}                                          

Here, for forming the expression for the WP reflected from the
boundary, we select the increasing part of stationary solution WF
$\alpha^{0} \exp(\xi x)$ only. Imposing the condition of continuity
on the time-dependent WF and its derivative at point $x = a$, we
obtain 2 new equations, from which we find the unknowns coefficients
$A_{T}^{0}$ and $\alpha^{0}$.

At the third step the WP, tunneling in region II, is incident upon
the first boundary of the barrier. Then it transforms into the WP,
transmitted through this boundary and propagated further in region
I, and into the WP, reflected from boundary and tunneled back in
region II. For WPs one can write:
\begin{equation}
\begin{array}{lcll}
\psi_{inc}^{3}(x, t) & = & \psi_{tr}^{2}(x, t), &
        \ 0 < x < a, \\
\psi_{tr}^{3}(x, t) & = & \int\limits_{0}^{+ \infty}
        G(E - \bar{E}) * A_{R}^{1}e^{-ikx -iEt/\hbar}dE, &
        \ x < 0, \\
\psi_{ref}^{3}(x, t) & = & \int\limits_{0}^{+ \infty}
        G(E - \bar{E}) * \beta^{1}e^{-\xi x -iEt/\hbar}dE, &
        \ 0 < x < a.
\end{array}
\end{equation}                                          

Using the conditions of continuity for the time-dependent WF and
its derivative at point $x = 0$, we obtain the unknowns coefficients
$A_{R}^{1}$ and $\beta^{1}$.

Analyzing further possible processes of transition (and reflection)
of WPs through the boundaries of barrier, we come to a deduction,
that any of the following steps can be reduced to one of 3 considered
above. For unknown coefficients $\alpha^{i}$, $\beta^{i}$,$A_{T}^{i}$
and $A_{R}^{i}$, used in expressions for WPs, forming in result of
some internal reflection from the boundaries, one can obtain the
recurrence relations.


Considering the propagation of WP by such way, we obtain the
expressions for WF on each region which can be written through
the series of multiple WPs. We determine the resultant
expressions for the incident, transmitted and reflected WPs in
relation to the barrier:
\begin{equation}
\begin{array}{lcll}
\psi_{inc}(x, t) & = & \int\limits_{0}^{+ \infty}
        G(E - \bar{E}) * e^{ikx -iEt/\hbar}dE, & x < 0, \\
\psi_{tr}(x, t) & = & \int\limits_{0}^{+ \infty}
        G(E - \bar{E}) * \sum\limits_{i=0}^{+ \infty}
        A_{T}^{i}e^{ikx - iEt/\hbar}dE, & x > a, \\
\psi_{ref}(x, t) & = & \int\limits_{0}^{+ \infty}
        G(E - \bar{E}) * \sum\limits_{i=0}^{+ \infty}
        A_{R}^{i}e^{-ikx - iEt/\hbar}dE, & x < 0.
\end{array}
\end{equation}                                          

The series of coefficients $\alpha^{n}$, $\beta^{n}$, $A_{T}^{n}$
and $A_{R}^{n}$ can be calculated on the basis of obtained reccurence
relations and have the following form:
\begin{equation}
\begin{array}{lcllcl}
\sum\limits_{n=0}^{+ \infty}A_{T}^{n}  & = &
        \displaystyle\frac{i4k \xi e^{-\xi a-ika}}{F_{sub}},    &
\sum\limits_{n=0}^{+ \infty}\alpha^{n} & = &
        \displaystyle\frac{2k (i\xi - k) e^{-2\xi a}}{F_{sub}}, \\
\sum\limits_{n=0}^{+ \infty}A_{R}^{n}  & = &
        \displaystyle\frac{k_{0}^{2}*D_{-}}{F_{sub}},           &
\sum\limits_{n=0}^{+ \infty}\beta^{n}  & = &
        \displaystyle\frac{2k (i\xi + k)}{F_{sub}},
\end{array}
\end{equation}                                          
where
\begin{equation}
\begin{array}{lll}
F_{sub}  & = & (k^{2}- \xi^{2}) D_{-}+ 2ik\xi D_{+}, \\
D_{\pm}  & = & 1 \pm e^{-2\xi a}, \\
k_{0}^{2}& = & k^{2}+ \xi^{2}= \displaystyle\frac{2mV_{1}}
{\hbar^{2}}.
\end{array}
\end{equation}                                          

All obtained expressions for the series $\sum \alpha^{n}$,
$\sum \beta^{n}$, $\sum A_{T}^{n}$ and $\sum A_{R}^{n}$ coincide
with the corresponding coefficients $\alpha$, $\beta$, $A_{T}$
and $A_{R}$ of the expressions (1), calculated by the stationary
methods \cite{Olkhovsky.1992.PRPLC,Razavy.1988.PRPLC}. Using
following substitution
\begin{equation}
i\xi \to k_{2},
\end{equation}                                          
where $k_{2}= \frac{1}{\hbar}\sqrt{2m(E - V_{1})}$ is the wave number
for the case of above-barrier energies, expression for coefficients
$\alpha^{n}$, $\beta^{n}$, $A_{T}^{n}$ and $A_{R}^{n}$ for each step,
expressions for WF for each step, the total WP transforms into
the corresponding expressions for the problem of the particle
propagation above this barrier.
Besides the following property is fulfilled:
\begin{equation}
\biggl |\sum\limits_{n=0}^{+ \infty}A_{T}^{n}\biggr |^{2}+
\biggl |\sum\limits_{n=0}^{+ \infty}A_{R}^{n}\biggr |^{2}= 1.
\end{equation}                                          



\section{Tunneling and reflecting times}

Let's find the time of leaving of WP from the barrier formed
in result of $n$ reflections from the boundaries (we name such WP
as $n$-fold packet). Using the expressions (4) for the first step,
one can determine the equation describing the propagation of the
maximum (peak) for incident, transmitted and reflected WPs in
relation to the first boundary:
\begin{equation}
\displaystyle\frac{\partial}{\partial E}arg \psi_{inc}(x, t) =
\displaystyle\frac{\partial}{\partial E}arg \psi_{tr}(x, t) =
\displaystyle\frac{\partial}{\partial E}arg \psi_{ref}(x, t) =
const.
\end{equation}                                          

Let the WP in region I be incident upon the first boundary of the
barrier in the initial time moment $t_{inc}$. One can find the time
$t_{ref}^{1}$ of the leaving of the reflected WP from the first
boundary to region I:
\begin{equation}
\begin{array}{lcl}
t_{ref}^{1} & = &
        t_{inc}+ \hbar\displaystyle\frac{\partial arg A_{R}^{0}}
        {\partial E}=
        t_{inc}+ \displaystyle\frac{2m}{\hbar k\xi}.
\end{array}
\end{equation}                                          

Similarly, for the time $t_{tr}^{1}$ of the leaving of the transmitted
WP from this boundary to region II we receive:
\begin{equation}
\begin{array}{lcl}
t_{tr}^{1} & = & t_{inc}+ \hbar\displaystyle\frac{\partial
        arg \beta^{0}}{\partial E}=
        t_{inc}+ \displaystyle\frac{m}{\hbar k\xi}.
\end{array}
\end{equation}                                          

Further we consider the second step, when the WP transits
through the second boundary of the barrier at point $x = a$.
Using the equations (5) and (12), one can obtain the time of
the leaving of the transmitted WP to region III and reflected
WP to region II from the second boundary:
\begin{equation}
\begin{array}{lcl}
t_{tr}^{2}  & = & t_{inc}+ \displaystyle\frac{ma}{\hbar k}+
                            \hbar\displaystyle\frac{\partial
                            arg A_{T}^{0}}{\partial E}=
                  t_{inc}+ \displaystyle\frac{2m}{\hbar k\xi}, \\
t_{ref}^{2} & = & t_{inc}+ \hbar\displaystyle\frac{\partial
                            arg \alpha^{0}}{\partial E}=
                  t_{inc}+ \displaystyle\frac{3m}{\hbar k\xi}.
\end{array}
\end{equation}                                  

Using this approach further, we find the time of the leaving of
the $n$-fold WP from the second boundary to region III for the
even step $n$:
\begin{equation}
\begin{array}{lcl}
t_{tr}^{n} & = & t_{inc}+ \displaystyle\frac{ma}{\hbar k}+
                       \hbar\displaystyle\frac{\partial
                       arg A_{T}^{n}}{\partial E}=
             t_{inc}+ (2n+1) \displaystyle\frac{2m}{\hbar k\xi}.
\end{array}
\end{equation}                                  

Similarly, for the odd step $n$ (except for first step) we find the
time of an leaving of the $n$-fold WP from the first boundary
to region I:
\begin{equation}
t_{ref}^{n}= t_{inc}+ \hbar\displaystyle\frac{\partial
                        arg A_{R}^{n}}{\partial E}=
              t_{inc}+ (n+1) \displaystyle\frac{4m}{\hbar k\xi}.
\end{equation}                                  



The complete WF, describing a particle transmitted through
barrier, represents total WP, the coordinates $(x, t)$ of
maximum of which can be identified with coordinates of a particle
(its most probable position), transmitted through the barrier and
propagated further in region III. The time corresponding to the
leaving of this maximum one can identify with the phase tunneling
time of particle through the barrier. One can obtain:
\begin{equation}
\begin{array}{lcl}
\tau_{tun}^{Ph} & = & t_{tr} - t_{inc}=
        \displaystyle\frac{ma}{\hbar k}+
        \hbar\displaystyle\frac{\partial arg \sum\limits_{n=0}^{+ \infty}
        A_{T}^{n}}{\partial E}= \\
& = &   \displaystyle\frac{m}{\hbar k\xi}*
        \displaystyle\frac{k_{0}^{4}sh (2\xi a) +
        2a\xi k^{2}(\xi^{2}-k^{2})}{4k^{2}\xi^{2}+ k_{0}^{4}sh^{2}(\xi a)}.
\end{array}
\end{equation}                                          

The phase time of reflection of a particle from the barrier takes into
account only all WPs leaving from the barrier in region I:
\begin{equation}
\tau_{ref}^{Ph}= t_{ref}- t_{inc}=
        \hbar\displaystyle\frac{\partial arg \sum\limits_{n=0}^{+ \infty}
        A_{R}^{n}}{\partial E}= \tau_{tun}^{Ph}.
\end{equation}                                          

Considering high enough (and wide enough) barrier (limit: $\xi a \gg 1$),
one can obtain the following expression for the phase times of tunneling
and reflection \cite{Olkhovsky.1992.PRPLC}:
\begin{equation}
\tau_{tun}^{Ph}= \tau_{ref}^{Ph}\to \displaystyle\frac{2}{v\xi},
\end{equation}                                          
where $v = \hbar k/m$ is the group velocity. One can see, that these
times at such limit coincide with time of reflection $t_{ref}^{1}$
at the first step and with time of transition $t_{tr}^{2}$ at the
second step. Therefore, considering $n$-fold transmitted and reflected
WPs at such limit the maximum of total WP is determined by first
packets only.

The phase times of tunneling and reflection, calculated using of
the method of multiple internal reflections, convergent with results
of \cite{Olkhovsky.1992.PRPLC}.

\section{The particle tunnels through the 1D- double-humb
rectangular barrier}

The method of multiple internal reflections allows simply enough to
calculate the amplitudes of transmitted and reflected WPs in relation
to the barrier for normalized incident WP, if the general stationary
solutions of WF are known. This method is effective in finding
the amplitudes for stationary WF. The solutions obtained by this
method for the problem with barriers of more composite form, than
rectangular, appear easier, than solutions obtained by stationary
methods.

\begin{figure}[p]
\centerline{\includegraphics[width=7cm]{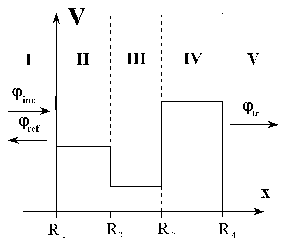}}
\caption{An one-dimension double-humb rectangular barrier}
\label{fig.2}
\end{figure}

Let's consider the particle propagating above the one-dimensional
double-humb rectangular barrier (see Fig.~\ref{fig.2}). Solving the
stationary Schr\"{o}dinger equations in each region, one can obtain
the general solution for stationary WF:
\begin{equation}
\varphi(x) = \left \{
\begin{array}{ll}
   e^{ikx}+A_{R}e^{-ikx}, & \mbox{for } x < R_{1}, \\
   \alpha_{2}e^{-ik_{2}x}+ \beta_{2}e^{ik_{2}x}, &
                                        \mbox{for } R_{1}< x < R_{2}, \\
   \alpha_{3}e^{-ik_{3}x}+ \beta_{3}e^{ik_{3}x}, &
                                        \mbox{for } R_{2}< x < R_{3}, \\
   \alpha_{4}e^{-ik_{4}x}+ \beta_{4}e^{ik_{4}x}, &
                                        \mbox{for } R_{3}< x < R_{4}, \\
   A_{T}e^{ikx}, & \mbox{for } x > R_{4},
\end{array}\right.
\end{equation}                                          
where $k = \frac{1}{\hbar}\sqrt{2mE}$,
    $k_{i}= \frac{1}{\hbar}\sqrt{2m (E-V_{i})}$ is the wave numbers,
    $i$ is the number of region. For finding the unknown coefficients
$A_{T}$, $A_{R}$, $\alpha_{i}$ and $\beta_{i}$ one can impose the
continuity condition for WF and its derivative at points $x = R_{i}$.
In result one can obtain 8 equations, from which the unknowns coefficients
can be calculated. In this the standard stationary solution method consists.

Let's find these coefficients, using the method of multiple internal
reflections for solution of this problem.
Analysing 7 undependent steps of WP propagation through the barrier
boundaries and finding the reccurent relations between unknown
koefficients $A_{T}^{i}$, $A_{R}^{i}$, $\alpha_{j}^{i}$ and
$\beta_{j}^{i}$, we can write the expressions for incident,
transmitted and reflected WPs in relation to the barrier. The sums
of coefficients can be calculated on the basis of obtained reccurent
relations and have the following form:
\begin{equation}
\begin{array}{lcl}
A_{R} & \to & \sum\limits_{n=0}^{+ \infty}A_{R}^{n}=
        R_{1}^{+}+ \displaystyle\frac{T_{1}^{+}Fr_{2}^{+}T_{1}^{-}}{1 -
        R_{1}^{-}Fr_{2}^{+}}, \\
\beta_{2} & \to & \sum\limits_{n=0}^{+ \infty}\beta_{2}^{n}=
        \displaystyle\frac{T_{1}^{+}}{1 - R_{1}^{-}Fr_{2}^{+}}, \\
\beta_{3} & \to & \sum\limits_{n=0}^{+ \infty}\beta_{3}^{n}=
        \displaystyle\frac{Ft_{2}^{+}}{1 - Fr_{2}^{-}Fr_{3}^{+}}, \\
\beta_{4} & \to & \sum\limits_{n=0}^{+ \infty}\beta_{4}^{n}=
        \displaystyle\frac{Ft_{3}^{+}}{1 - Fr_{3}^{-}R_{4}^{+}}, \\
A_{T} & \to & \sum\limits_{n=0}^{+ \infty}A_{T}^{n}=
        T_{4}^{+}* \beta_{4},
\end{array}
\end{equation}                                          
where
\begin{equation}
\begin{array}{lcllcl}
Fr_{2}^{+} & = &
        R_{2}^{+}+ \displaystyle\frac{T_{2}^{+}Fr_{3}^{+}T_{2}^{-}}
       {1 - Fr_{3}^{+}R_{2}^{-}}, &
Fr_{3}^{+} & = &
        R_{3}^{+}+ \displaystyle\frac{T_{3}^{+}R_{4}^{+}T_{3}^{-}}
       {1 - R_{4}^{+}R_{3}^{-}}, \\
Fr_{2}^{-} & = &
        R_{2}^{-}+ \displaystyle\frac{T_{2}^{-}R_{1}^{-}T_{2}^{+}}
       {1 - R_{2}^{+}R_{1}^{-}}, &
Fr_{3}^{-} & = &
        R_{3}^{-}+ \displaystyle\frac{T_{3}^{-}Fr_{2}^{-}T_{3}^{+}}
       {1 - R_{3}^{+}Fr_{2}^{-}}, \\
Ft_{2}^{+} & = &
        \displaystyle\frac{T_{1}^{+}T_{2}^{+}}
       {1 - R_{2}^{+}R_{1}^{-}}, &
Ft_{3}^{+} & = &
        \displaystyle\frac{Ft_{2}^{+}T_{3}^{+}}
       {1 - R_{3}^{+}Fr_{2}^{-}}.
\end{array}
\end{equation}                                          

The coefficients $T_{i}^{\pm}$ and $R_{i}^{\pm}$ have the following
form:
\begin{equation}
\begin{array}{lcllcl}
T_{1}^{+}& = & \displaystyle\frac{2k}{k + k_{2}}
                e^{i (k-k_{2}) R_{1}}, &
T_{1}^{-}& = & \displaystyle\frac{2k_{2}}{k + k_{2}}
                e^{-i (k-k_{2}) R_{1}}, \\
T_{2}^{+}& = & \displaystyle\frac{2k_{2}}{k_{2}+ k_{3}}
                e^{i (k_{2}-k_{3}) R_{2}}, &
T_{2}^{-}& = & \displaystyle\frac{2k_{3}}{k_{2}+ k_{3}}
                e^{-i (k_{2}-k_{3}) R_{2}}, \\
T_{3}^{+}& = & \displaystyle\frac{2k_{3}}{k_{3}+ k_{4}}
                e^{i (k_{3}-k_{4}) R_{3}}, &
T_{3}^{-}& = & \displaystyle\frac{2k_{4}}{k_{3}+ k_{4}}
                e^{-i (k_{3}-k_{4}) R_{3}}, \\
T_{4}^{+}& = & \displaystyle\frac{2k_{4}}{k_{4}+ k}
                e^{i (k-k_{4}) R_{4}}, &
T_{4}^{-}& = & \displaystyle\frac{2k}{k_{4}+ k}
                e^{-i (k-k_{4}) R_{4}},
\end{array}
\end{equation}                                          
\begin{equation}
\begin{array}{lcllcl}
R_{1}^{+}& = & \displaystyle\frac{k-k_{2}}{k + k_{2}}
                e^{2ikR_{1}}, &
R_{1}^{-}& = & \displaystyle\frac{k_{2}-k}{k_{2}+ k}
                e^{-2ik_{2}R_{1}}, \\
R_{2}^{+}& = & \displaystyle\frac{k_{2}-k_{3}}{k_{2}+ k_{3}}
                e^{2ik_{2}R_{2}}, &
R_{2}^{-}& = & \displaystyle\frac{k_{3}-k_{2}}{k_{3}+ k_{2}}
                e^{-2ik_{3}R_{2}}, \\
R_{3}^{+}& = & \displaystyle\frac{k_{3}-k_{4}}{k_{3}+ k_{4}}
                e^{2ik_{3}R_{3}}, &
R_{3}^{-}& = & \displaystyle\frac{k_{4}-k_{3}}{k_{4}+ k_{3}}
                e^{-2ik_{4}R_{3}}, \\
R_{4}^{+}& = & \displaystyle\frac{k_{4}-k}{k_{4}+ k}
                e^{2ik_{4}R_{4}}, &
R_{4}^{-}& = & \displaystyle\frac{k-k_{4}}{k + k_{4}}
                e^{-2ikR_{4}}.
\end{array}
\end{equation}                                          

From these expressions using the substitution similar (10), one can
obtain the corresponding solutions for the problem when the particle
tunnels under the barrier having the same form.

Applying the method of multiple internal reflections by this way, one
can find the expressions for WPs (and also amplitudes for
corresponding stationary WF) in the above-barrier regions for an
one-dimensional problem with a barrier consisting of an arbitrary
finite number of potential steps. To receive such solutions using the
usual stationary approach it appears much more complicatedly. This
perspective of the method of multiple internal reflections is found
out for the first time and is displayed the more brightly, if the
barrier have more composite form.


\section{The scattering of a particle on spherically symmetric
rectangular barrier}

\begin{figure}[p]
\centerline{\includegraphics[width=7cm]{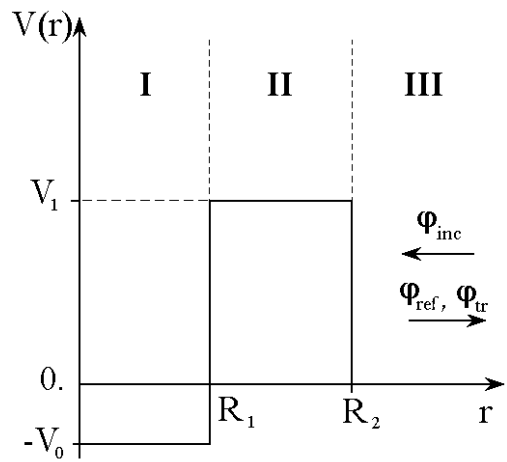}}
\caption{A spherically symmetric rectangular barrier}
\label{fig.3}
\end{figure}

Let's consider the problem of scattering of a particle on the
spherically symmetric field, the radial part of which looks like
(see Fig.~\ref{fig.3}):
\begin{equation}
V (r) = \left \{
\begin{array}{rll}
  -V_{0}, & \mbox{for } r < R_{1};        & \mbox{(region I)}; \\
   V_{1}, & \mbox{for } R_{1}< r < R_{2}, & \mbox{(region II)}; \\
   0,     & \mbox{for } r > R_{2},        & \mbox{(region III)}.
\end{array}\right.
\end{equation}                                          

We study the case, when the moment $l = 0$ and the energy levels
lay below than the height of barrier. Let's consider the standard
stationary method of problem
solution. Solving the stationary Schr\"{o}dinger equation in each
region, we obtain the general solution of WF:
\begin{equation}
\psi (r, \theta, \varphi) = \frac{\chi (r)}{r}
    * Y_{lm}(\theta, \varphi),
\end{equation}                                          
\begin{equation}
\chi (r) = \left \{
\begin{array}{lll}
A (e^{-ik_{1}r}- e^{ik_{1}r}), & \mbox{for }r < R_{1}, &
        \mbox{(region I)}, \\
\alpha e^{\xi r}+ \beta e^{-\xi r}, & \mbox{for }R_{1}< r < R_{2}, &
        \mbox{(region II)}, \\
e^{-ikr}+ Se^{ikr}, & \mbox{for }r > R_{2}; & \mbox{(region III)},
\end{array}\right.
\end{equation}                                          
where $Y_{lm}(\theta, \varphi)$ is the spherical function,
    $k_{1} = \frac{1}{\hbar}\sqrt{2m (E + V_{0})}$,
    $\xi = \frac{1}{\hbar}\sqrt{2m (V_{1} - E)}$,
    $k = \frac{1}{\hbar}\sqrt{2mE}$.
One can find the unknown coefficients $S$, $A$, $\alpha$ and $\beta$
from the continuity condition for WF and its derivative at points
$r = R_{1}$ and $r = R_{2}$. With the stationary point of view the
tunneling of particle can be described on the basis of spherical
vawes. In the obtained solution it is impossible to separate the
vawe transmitted through the barrier from wave reflected from the
barrier, and, therefore, to separate their amplitudes (only one
item $Se^{ikr}$ contains both transmitted, and reflected waves).

The method of multiple internal reflections allows to find the solution
of this problem. Let's apply it to this problem. We study the propagation
of WPs sequentially on steps of its transition in relation to each
of boundaries of the barrier (similarly to one-dimensional problem).

In result of multiple internal reflections (and transitions) in
relation to the boundaries of the barrier the total time-dependent
WF in each region can be written in form of series, composed
from the incoming and outcoming WPs. One can calculate the series:
\begin{equation}
\begin{array}{lcl}
\sum\limits_{n=1}^{+ \infty}S^{n}       & = &
        \displaystyle\frac{4ik\xi
        \biggl(\displaystyle\frac{i\xi-k_{1}}{i\xi+k_{1}}-
        e^{2ik_{1}R_{1}}\biggr) e^{-2\xi (R_{2}-R_{1}) - 2ikR_{2}}}
        {F_{sub} * (k+i\xi)^{2}}, \\
\sum\limits_{n=0}^{+ \infty}A^{n}       & = &
        \displaystyle\frac{4ik\xi e^{-ikR_{2}+ ik_{1}R_{1}-
        \xi (R_{2}-R_{1})}}{F_{sub}* (k+i\xi) (k_{1}+i\xi)}, \\
\sum\limits_{n=0}^{+ \infty}\alpha^{n}  & = &
        \displaystyle\frac{2k \biggl (1 + \displaystyle\frac{k_{1}-i\xi}
        {k_{1}+i\xi}e^{2ik_{1}R_{1}}\biggr) e^{- (\xi+ik) R_{2}}}
        {F_{sub}* (k+i\xi)}, \\
\sum\limits_{n=0}^{+ \infty}\beta^{n}   & = &
        \alpha^{0}* \displaystyle\frac{R_{1}^{-}(1 - R_{1}^{+}R_{0}^{-})
        + T_{1}^{-}R_{0}^{-}T_{1}^{+}}{F_{sub}},
\end{array}
\end{equation}                                          
where
\begin{equation}
\begin{array}{lcl}
F_{sub} & = &
        1 + \displaystyle\frac{k_{1}-i\xi}{k_{1}+i\xi}
        e^{2ik_{1}R_{1}} -
        \displaystyle\frac{(k-i\xi)(k_{1}-i\xi)}{(k+i\xi)(k_{1}+i\xi)}
        e^{-2\xi (R_{2}- R_{1})} -      \\
& - &   \displaystyle\frac{k-i\xi}{k+i\xi}
        e^{-2\xi (R_{2}-R_{1}) + 2ik_{1}R_{1}}.
\end{array}
\end{equation}                                          

Now we determine the incident, transmitted and reflected WPs in
relation to the barrier at whole. Considering them for region III,
we write:
\begin{equation}
\begin{array}{lcl}
\chi_{inc}(r, t) & = & \int\limits_{0}^{+ \infty}g (E - \bar{E})
        \theta (V_{1} - E) * e^{-ikr -iEt/\hbar}dE, \\
\chi_{tr}(r, t) & = & \int\limits_{0}^{+ \infty}g (E - \bar{E})
        \theta (V_{1} - E) * S_{tr}* e^{ikr -iEt/\hbar}dE, \\
\chi_{ref}(r, t) & = & \int\limits_{0}^{+ \infty}g (E - \bar{E})
        \theta (V_{1} - E) * S_{ref}e^{ikr -iEt/\hbar}dE,
\end{array}
\end{equation}                                          
where
\begin{equation}
\begin{array}{lcl}
S_{tr}  & = & \sum\limits_{n=1}^{+ \infty}S^{n},        \\
S_{ref} & = & S^{0},                                    \\
S       & = & S_{tr}+ S_{ref}.
\end{array}
\end{equation}                                          

The expression $S$ represents the S-matrix. Thus, using the method
of multiple internal reflections it appears possible to divide
it into two components corresponding to amplitudes of transmitted
and reflected WPs in relation to the barrier at whole. This
property having physical sense, is obtained for the first time.

Expressions for coefficients $S^{n}$, $A^{n}$, $\alpha^{n}$ and
$\beta^{n}$ for each step, expression for WF for each step,
the series of coefficients $S^{n}$, $A^{n}$, $\alpha^{n}$ and
$\beta^{n}$ under the substitution (10) transform into the
corresponding expressions for the solution of problem of WP
propagation above the barrier.
The series of coefficients $S^{n}$, $A^{n}$, $\alpha^{n}$ and
$\beta^{n}$ coincide with the corresponding coefficients $S$, $A$,
$\alpha$ and $\beta$, calculated by stationary methods.


Similarly to the equations (12) for the one-dimensional problem one can
determine the equation for propagation of the maximum of incident,
transmitted and reflected WPs in relation to the barrier for the
spherically symmetric problem. One can obtain the
times necessary for transitting WP through the barrier and for
reflecting WP from the barrier:
\begin{equation}
\begin{array}{lcl}
\tau_{tun}^{Ph}& = & t_{tr}- t_{inc}=
        \displaystyle\frac{2mR_{2}}{\hbar k}+
        \hbar\displaystyle\frac{\partial arg S_{tr}}{\partial E}, \\
\tau_{ref}^{Ph}& = & t_{ref}- t_{inc}=
        \displaystyle\frac{2mR_{2}}{\hbar k}+
        \hbar\displaystyle\frac{\partial arg S_{ref}}{\partial E}.
\end{array}
\end{equation}                                          

For the problem of particle tunneling under the barrier we receive:
\begin{equation}
\begin{array}{lcl}
\tau_{tun}^{Ph}& = & \hbar\displaystyle\frac{\partial}{\partial E}
        arg \displaystyle\frac{i\xi-k_{1}-(i\xi+k_{1})e^{2ik_{1}R_{1}}}
       {(i\xi+k)^{2}(i\xi+k_{1}) F_{sub}}, \\
\tau_{ref}^{Ph}& = & \displaystyle\frac{2m}{\hbar\xi k},
\end{array}
\end{equation}                                          
and for the problem of WP propagating above the barrier
we receive:
\begin{equation}
\begin{array}{lcl}
\tau_{tun}^{Ph}& = & \displaystyle\frac{2m(R_{2}-R_{1})}{\hbar k_{2}}+
         \hbar\displaystyle\frac{\partial}{\partial E}
        arg \displaystyle\frac{k_{2}-k_{1}-(k_{2}+k_{1})e^{2ik_{1}R_{1}}}
       {(k+k_{2}) (k_{1}+k_{2}) F_{above}}, \\
\tau_{ref}^{Ph}& = & 0,
\end{array}
\end{equation}                                          
where $F_{above}$ can be obtained from the expression for $F_{sub}$
using the substitution (10).

Let's consider the enough high and wide barrier (limits: $\xi (R_{2} -
R_{1}) \to \ + \infty$, $\xi \to + \infty$). For tunneling time we
obtain the following expression:
\begin{equation}
\tau_{tun}^{Ph}= \displaystyle\frac{2m}{\hbar k \xi}+
        \displaystyle\frac{4mR_{1}\sin{2k_{1}R_{1}}(1-2\cos{2k_{1}R_{1}})}
        {\hbar \xi (1-\cos{2k_{1}R_{1}})}.
\end{equation}                                          

The tunneling time does not depend on barrier width (the effect
of Hartman-Fletcher), but depends on $k_{1}$ and $R_{1}$.

\section{Conclusions}

At the first time the non-stationary method, describing a process
of tunneling of a particle through the barrier on the basis of
consideration of multiple internal reflections of wave packets
relatively with the barrier boundaries and named as \emph{the method
of multiple internal reflections}, is presented in this paper.
In accordence with this method one can describe the tunneling process
in dependence on time and study the specific features of tunneling
at any interesting moment of time and point of space in details.

The one-dimensional stationary problem of tunneling of the particle
through the rectangular barrier with taking into account of the
multiple internal reflections was earlier solved
\cite{McVoy.1967.RMP,Anderson.1989.AJPIA}. The new specific
perspective of the method of multiple internal reflections in solving
of this problem is a possibility
to fulfill the time analysis of tunneling and reflection in relation
to the barrier (with consideration of multiple internal reflections
of wave packets).

Using the method of multiple internal reflections the problem of
tunneling of a particle through spherically symmetric barrier is
considered at the first time and splved. Here, using this method
it is possible (as against the known stationary approaches) to
separate the wave packet transmitted through the barrier, from the
wave packet reflected from the barrier (both packets are spherically
divergent).
In result, one can calculate such stationary parameters as the
coefficient of penetrability of particle through the barrier and the
coefficient of reflectivity of particle from the barrier.

In result for the spherically symmetric problem for S-matrix the following
property is fulfilled:
\[
S = S_{tr}+ S_{ref},
\]
i.~e. it consists of two components corresponding to the transmission
and reflection of particle in relation to barrier. This property has
physical sense and is justified.

\newpage

\bibliographystyle{h-physrev4}
\bibliography{UPJ1}

\newpage
\listoffigures

\end{sloppypar}
\end{document}